# Construction and Usage of a Human Body Common Coordinate Framework Comprising Clinical, Semantic, and Spatial Ontologies

Katy Börner[1], Ellen M. Quardokus[1], Bruce W. Herr II[1], Leonard E. Cross[1], Elizabeth G. Record[1], Yingnan Ju[1], Andreas D. Bueckle[1], James P. Sluka[1], Jonathan C. Silverstein[2], Kristen M. Browne[3], Sanjay Jain[4], Clive H. Wasserfall[5], Marda L. Jorgensen[5], Jeffrey M. Spraggins[6], Nathan H. Patterson[6], Mark A. Musen,[7] Griffin M. Weber[8]

[1] Department of Intelligent Systems Engineering, Luddy School of Informatics, Computing, and Engineering, Indiana University, Bloomington, Indiana, USA
[2] Department of Biomedical Informatics, University of Pittsburgh School of Medicine, Pittsburgh, Pennsylvania, USA
[3] Department of Health and Human Services, National Institute of Allergy and Infectious Diseases, The National Institutes of Health, Bethesda, Maryland, USA
[4] Department of Medicine, Washington University School of Medicine, Saint Louis, Missouri, USA
[5] Departments of Pathology and Pediatrics, University of Florida, Gainesville, Florida, USA
[6] Mass Spectrometry Research Center, Vanderbilt University, Nashville, Tennessee, USA
[7] Stanford Center for Biomedical Informatics Research, Stanford University, Stanford, California, USA
[8] Department of Biomedical Informatics, Harvard Medical School, Boston, Massachusetts, USA

**Abstract**
The National Institutes of Health's (NIH) Human Biomolecular Atlas Program (HuBMAP) aims to create a comprehensive high-resolution atlas of all the cells in the healthy human body. Multiple laboratories across the United States are collecting tissue specimens from different organs of donors who vary in sex, age, and body size. Integrating and harmonizing the data derived from these samples and "mapping" them into a common three-dimensional (3D) space is a major challenge. The key to making this possible is a "Common Coordinate Framework" (CCF), which provides a semantically annotated, 3D reference system for the entire body. The CCF enables contributors to HuBMAP to "register" specimens and datasets within a common spatial reference system, and it supports a standardized way to query and "explore" data in a spatially and semantically explicit manner. This paper describes the construction and usage of a CCF for the human body and its reference implementation in HuBMAP. The CCF consists of (1) a CCF Clinical Ontology, which provides metadata about the specimen and donor (the "who"); (2) a CCF Semantic Ontology, which describes "what" part of the body a sample came from and details anatomical structures, cell types, and biomarkers (ASCT+B); and (3) a CCF Spatial Ontology, which indicates "where" a tissue sample is located in a 3D coordinate system. CCF design starts with domain experts manually selecting terms from existing ontologies and organizing them in so called ASCT+B tables that guide the development of the CCF Semantic Ontology. The ASCT+B tables also guide the design of the HuBMAP CCF Reference Object Library which was initially populated with 3D reference organ models obtained from the Visible Human project provided by the National Library of Medicine; the CCF Spatial Ontology represents the spatial size, position, and orientation of tissue data in relationship to the reference organ. The CCF Clinical Ontology is a proper subset of clinical metadata associated with the tissue samples that are relevant to CCF design. An initial version of all three CCF ontologies has been implemented for the first HuBMAP Portal release. It was successfully used by Tissue Mapping Centers to semantically annotate and spatially register 48 kidney and spleen tissue blocks. The blocks can be queried and explored in their clinical, semantic, and spatial context via the CCF user interface in the HuBMAP Portal.

**1. Introduction**
The Human BioMolecular Atlas Program (HuBMAP) is a large, multi-institutional project, funded by the National Institutes of Health to create a detailed spatial map of all the cells in the human body [1]. Similar to how the Human Genome Project identified all the base pairs of the human genome twenty years ago, HuBMAP aims to make a similar leap forward in our understanding of the organization and function of cells by using next generation tools for high-throughput imaging and omics assays to generate multi-modal 3D tissue maps down to single cell resolution.

Other projects are also working toward this goal, including NIH funded efforts such as LungMAP [2], (Re)building the Kidney (RBK) [3], Kidney Precision Medicine Project (KPMP) [4], Human Tumor Atlas Network (HTAN) [5], the genitourinary developmental molecular anatomy project (GUDMAP) [6], Stimulating Peripheral Activity to Relieve Conditions (SPARC), the Brain Research through Advancing Innovative Neurotechnologies (BRAIN) Initiative [7], as well as non-governmental funded efforts such as the Human Cell Atlas (HCA) funded by the Chan Zuckerberg Initiative [8–10], and the recently launched Helmsley Charitable Trust: Gut Cell Atlas (GCA).

While some projects focus on one organ, HuBMAP targets multiple organs. Laboratories across the United States are collecting tissue specimens from numerous donors who vary in sex, age, and body size among other attributes. Data range from volumetric imaging data (CT or MRI scans) to spatially resolved single-cell biomolecular data derived from a wide array of technologies and methods including transcriptomic, proteomic, lipidomic, and metabolomic studies. A major challenge for HuBMAP is harmonizing these different data sources and aligning them in a common, semantically annotated 3D space. The key to making this possible is a "Common Coordinate Framework" (CCF). This paper describes the construction and usage of a CCF for the human body and its implementation in the HuBMAP Portal.

The 2017 NIH Common Coordinate Framework meeting coined the term CCF and defined it as a coordinate system that uniquely and reproducibly defines any location in the human body [11]. Papatheodorou described it as a "spatiotemporal computational framework for the management, integration, and analysis of anatomically and spatially indexed data" [12]. Rood et al. [13] define a CCF as "an underlying reference map of organs, tissues, or cells that allows new individual samples to be mapped to determine the relative location of structural regions between samples."

The working definition we use for this paper is as follows: A CCF consists of ontologies and reference object libraries, computer software (e.g., user interfaces), and training materials that support the efficient mapping, registration, and exploration of clinically, semantically, and spatially indexed human tissue data. The HuBMAP CCF consists of (1) a CCF Clinical Ontology, which provides CCF relevant demographic and clinical metadata about the specimen and donor (the "who"); (2) a CCF Semantic Ontology, which describes "what" part of the body a tissue sample came from; and (3) a CCF Spatial Ontology, which indicates "where" the tissue is located in a 3D reference system (RS). In addition, the CCF contains a "registration process" (RP) that makes it possible to annotate data and map them to the RS, and an "exploration process" (EP), which facilitates query, analysis, and visual examination of registered data.

## 2. Materials and Methods
This section discusses requirements for a CCF as motivated by HuBMAP; we then present the CCF Knowledge Architecture that defines the data structures that define and interlink the clinical, semantic, and spatial ontologies; and then discuss and exemplify all three ontologies.

### 2.1. Requirements for the HuBMAP CCF
There are different approaches to developing a CCF. We were primarily guided by the objectives of HuBMAP (described below); however, our CCF ultimately needs to be generalizable to other applications and designed in a way that HuBMAP can collaborate with other efforts to map the human body.

For the initial CCF presented in this paper, we aimed to address the following three core requirements:
1) The CCF must support two general types of research questions for HuBMAP relevant for "mapping": What are the spatial positions and/or distributions of all the cell types in a given anatomical structure; and, what are the spatial positions and/or distributions of all the anatomical structures that contain a given cell type?
2) The CCF must be able to "register" 2D images of tissue sections and 3D volumes of tissue blocks within a well-defined 3D reference system (RS) using a registration process (RP). Registration includes spatial positioning and semantic annotation of tissue samples.
3) The CCF must support the "exploration" of tissue datasets based on clinical data including donor patient characteristics (e.g., age, sex, ethnicity, BMI) and data type and origin; based on semantic annotations of anatomical structures, cell types, and biomarkers (e.g., gene, protein, lipid, or metabolic markers); but also based on absolute or relative spatial location in the body.

In general, the CCF must work at all length scales, from the entire human body (meter) to macro (centimeter) and micro anatomy (millimeter) to single cells (micrometer).

## 2.2. CCF Knowledge Architecture

The CCF Knowledge Architecture consists of three components: a CCF Clinical Ontology, a CCF Semantic Ontology, and a CCF Spatial Ontology (Figure 1). These have been defined as a formal ontology using semantic web technologies in Web Ontology Language (OWL 2) [14]. Data is added as RDF/XML, or JSON-LD that is translated to RDF/XML. As such, it is compatible with, and can be linked to, other ontologies easily. We have deposited these three ontologies in BioPortal, see https://bioportal.bioontology.org/ontologies/CCF.

### 2.2.1. CCF Clinical Ontology

The CCF Clinical Ontology describes the demographic and clinical, workflow, and other metadata associated with human tissue samples. The complete HuBMAP clinical data—covering more than 100 metadata fields—was reduced to a smaller set of 21 metadata fields that is relevant for CCF design and usage. The current CCF subset includes demographics and clinical data (e.g., sex, age, BMI), workflow information (e.g., tissue sample creation/modification date, donor/organ/tissue ID, specimen/data/assay type), and author information (e.g., author group/creator). All data is stored in a Neo4J graph database, which can be exported in W3C Prov format that was developed to support the interchange of provenance information on the Web [15]. Figure 1a shows major CCF Clinical Ontology classes, which include information about the donor (demographics and clinical data), tissue sample, and derived datasets. Additional components of the CCF Clinical Ontology link samples to the laboratories that collected the tissue and methods used. While important for HuBMAP, these are outside of the scope of the initial CCF release and this paper.

**2.2.2. CCF Semantic Ontology**

The CCF Semantic Ontology lists the names of anatomical structures and cell types (ASCT) and their *part_of* relationship to each other ("partonomy"). For example, as illustrated in Figure 2a, podocyte cells are part of the glomerulus, which is part of the nephron, a functional tissue unit (FTU) of the kidney. To simplify CCF design and usage, we adopt a 'nested objects' view of the human body, where each anatomical structure and cell can only be part of one higher-level object. All structures are disjoint, but several can *touch* each other. Certain structures, like blood vessels, might *surround* or *pass through* others. Each anatomical structure consists of different cell types. The same cell types might exist in multiple organs; however, each individual cell has exactly one location.

The CCF Semantic Ontology has a class *Entity* (e.g., a tissue block or tissue section) that is connected to the Sample in the CCF Clinical Ontology and is annotated by one or more *Ontology Terms* (Figure 1b). There are four main properties: *ccf_annotation* maps an Entity to an ASCT *Ontology Term*; *ccf_freetext_annotation* provides additional details about an entity as free text (allowing annotations that cannot be easily mapped to existing ontologies); *ccf_same_as* associates *ccf_freetext_annotation* annotations or external terms (i.e., from another ontology) with an ASCT term (e.g., after terminology differences are resolved); and, *ccf_part_of* indicates the hierarchy of nested objects. Note that some *ccf_part_of* relationships are covered in existing ontologies while others are not.

CCF Semantic Ontology design starts by working with organ experts to manually construct "ASCT tables", which capture HuBMAP-relevant partonomies of anatomical structures (AS) and the cell types (CT) present in the AS (Figure 2b). The tables also list major biomarkers (e.g., cell type-specific gene, protein, lipid, and metabolite expression profiles), resulting in an ASCT+B table. The tables are built as spreadsheets and are then converted into OWL [16]. Next, we identify ASCTs names and unique identifiers in existing ontologies, such as Foundational Model of Anatomy (FMA) [17,18], UBERON [19], and Cell Ontology [20]. The existing ontologies have tens of thousands of terms, many of which are out of scope of HuBMAP's focus on healthy human adults; examples are concepts for capturing development and growth, cross-species comparisons, and disease. The CCF Semantic Ontology is much smaller. For example, the ASCT+B for the kidney has 39 anatomical structures, 54 cell types, and 81 biomarkers while the spleen features 33 anatomical structures, 23 cell types, and 42 biomarkers. These subsets can be expanded in the future to cover new HuBMAP data and use cases.

**2.2.3. CCF Spatial Ontology**

The CCF Spatial Ontology describes the 2D and 3D shapes of entities and their physical locations and orientations (Figure 2a). It consists of three main classes (Figure 1c):
- A *Spatial Entity* defines a bounded Cartesian space and its measurement units. It typically represents a real-world thing, e.g., a human body, a human kidney, a tissue section, or an individual cell. By using the *ccf_representation_of* property, we say that a *Spatial Entity* is representing/standing in for either an ASCT term in the CCF Semantic Ontology or a physical object, such as a tissue sample. *Spatial Entities* connect to ASCT+B terms using either *ccf_representation_of* or *ccf_annotation*.
- A *Spatial Object Reference* provides a reference to an external representation of a *Spatial Entity*, such as a 3D object file (e.g., in obj, fbx, gltf format) or a 2D image (e.g., in tiff, png, svg format).
- A *Spatial Placement* defines how to place a *Spatial Entity* or *Spatial Object Reference* relative to another *Spatial Entity*, using scaling, rotation, and translation (in that order). Note that rotation (in x, y, z order) occurs around the center of the object's coordinate space; by default, rotation is considered in Euler order. In the case of *Spatial Object References*, it defines how to transform a 2D or 3D object so that it fits the *Spatial Entity*'s dimensions and units. In the case of *Spatial Entities*, it shows how to place one *Spatial Entity* relative to another.

The *Spatial Object Reference* points to an Object in the CCF Reference Object Library that is initially populated with anatomically correct 3D reference organs created using male and female data from the Visible Human Project made available by the National Library of Medicine [21-22]. To better reflect the range of human diversity, we are in the process of developing consensus organs using magnetic resonance imaging (MRI), computed tomography (CT) scans, and anatomical images from hundreds of male and female donors.

## 3. Results
This section discusses our current ASCT+B tables and their usage for registration and exploration of tissue blocks and sections.

### 3.1. Anatomical Structures, Cell Types, and Biomarkers Tables
As of June, 2020, more than 30 domain experts from different consortia—including HCA, SPARC, KPMP, RBK, LungMAP, GUDMAP, HTAN, BRAIN Initiative Cell Census Network (BICCN), and the Allen Brain Institute—have constructed draft ASCT+B tables for eight organs: kidney, spleen, lymph nodes, heart, liver, skin, small and large intestine. The domain experts bring expertise in anatomy and pathology, immunology, genetics, and proteomics. Each table has an average of 26 unique anatomical structures (range 17-78), 29 cell types (range 16-54), and 61 biomarkers (range 37-83). Simplified views of the AS and CT portions of the kidney and spleen tables are shown in Supplements S1-S4, with mappings to Uberon ontology IDs. The HuBMAP CCF Ontology source code repository is available at http://purl.org/ccf/source.

### 3.2. Spatial Ontology
As of June 2020, the Object Library contains two reference organs (left and right kidneys and spleens) from the Visible Human male and female dataset [22] for a total of six 3D nested organ objects (Figure 3 and Supplement S5). The male dataset comprises 1,871 cross-sections at 1mm intervals for both CT and anatomical images at a resolution of 4,096 pixels by 2,700 pixels. The female data set has the same characteristics as the Visible Human male but axial anatomical images were obtained at 0.33 mm intervals resulting in 5,189 cross-section anatomical images. The male was white, 180.3 cm (71 inch) tall, 199 pounds and was 38 years old when he died in 1993. The female was white, 171.2 cm (67.4 inch) tall, obese (weight not available), and 59 years old when she died.

HuBMAP is currently constructing a consensus reference kidney based on 250 female and 250 male individuals. Key patient demographics such as sex, race, ethnicity, age, weight, height and BMI are collected. The consensus kidney, called VU500-kidney, uses 3D abdomen micro-CT images (~1mm isotropic resolution) available through ImageVU, a Vanderbilt University Medical Center database of MR and CT imaging data that are linked to de-identified clinical metadata. Custom multi-atlas registration and segmentation pipelines [23–26] are employed to create average and variability maps with ~1 mm isotropic resolution across 500 individuals' organs. Data comes as pixel volume. Segmentation is used to compile 3D reference objects. Current resolution is sufficient for extracting the outer shape of organs but insufficient for extracting inner anatomical structures (e.g., cortex, medullary pyramids, or calyces).

For the first HuBMAP Portal Release, two reference organs (kidney and spleen) are made freely available in GLB format, a binary form of the nested Graphics Library Transmission Format (glTF) developed by the Khronos Group 3D Formats Working Group [27]. The 3D reference object files can be used for API-neutral runtime asset delivery of 3D scenes and models using the JSON standard. Objects can be viewed and explored using free web browsers, e.g., Babylon.js [28]. Screenshots and major properties of the six nested organ objects are given in Table 1.

The 3D reference objects used in the Spatial Ontology might use alternative naming schemas than those in the Semantic Ontology, in which case mapping tables are provided. All reference objects are available at http://purl.org/ccf/source/objects and basic properties are provided in the CCF Portal [29].

### 3.3. Implementation and Usage

The HuBMAP CCF is used by data providers to spatially register and semantically annotate data, and by HuBMAP portal users to search, filter, and explore data. We developed two software tools, which leverage the CCF, to assist in these "registration" and "exploration" processes (Supplement S6). We describe them briefly here; though, details of the software architecture and design are outside the scope of this paper.

Registration Process: Registration User Interface (RUI): Tissue extraction locations are typically documented using photographs or videos and rarely capture the precise size, position, or rotation of these tissue blocks. The RUI addresses this by providing a graphical method that enables users to document the tissue extraction site, in relation to the donor organ, by drag-and-drop positioning a correctly sized tissue block inside a reference organ pulled from the 3D Reference Object Library. The RUI requires about 5 minutes of training time and 2 minutes for each tissue registration. To date, it has been used to register 48 tissue blocks (Supplement S7). The RUI for the kidney can be explored at https://hubmapconsortium.github.io/ccf-3d-registration/.

Exploration Process: Exploration User Interface (EUI): The EUI enables users to explore 2D and 3D tissue data both semantically and spatially across multiple scales using the HuBMAP CCF. Through a split-screen interface, users can navigate HuBMAP data, while seeing "where" they are in both the semantic ontology partonomy (e.g., "kidney : cortex") and within an anatomically correct 3D reference object. The EUI uses semantic annotations to support search, browsing, and filtering. The EUI can be explored via the HuBMAP Portal at https://portal.hubmapconsortium.org; login and select CCF in top-right of the navigation menu.

### 3.4. Instantiation of the CCF Ontologies

As illustrated in Figure 4, when a tissue sample, such as a kidney specimen, are registered and annotated through the RUI, they are assigned unique identifiers (e.g., "UUID-S-5678" linked to "Donor UUID-D-1234") using the CCF Clinical Ontology shown on top. They are linked to a term in the CCF Semantic Ontology given in lower-left, indicating the anatomical structure or cell type (e.g., "kidney cortex"). The CCF Semantic Ontology's anatomical structures partonomy shows how the sample fits within larger structures, up to the whole body. Using the CCF Spatial Ontology shown in lower-right, samples are also linked to a *Spatial Entity* (e.g., "UUID-SE-9123"), which gives its size/dimensions. A *Spatial Placement* (e.g., "UUID-SP-4567") positions the sample relative to another *Spatial Entity* (e.g., "#VHKidney").

### 3.5. Initial Validation

The CCF reference objects and all CCF ontologies presented in this paper have been examined and approved by organ experts. The accuracy and reproducibility of tissue block registrations using the RUI is under examination via a separate human subject study [30]. That study will also capture information on the time it takes to register a tissue using different user interfaces. A user study that examines task accuracy and completion time for different exploration tasks using the RUI is in progress.

### 4. Discussion

In this paper, we introduced the CCF we are developing for HuBMAP. The CCF meets the three core requirements discussed in Section 2.1. The CCF is a work-in-progress, with CCF Semantic and Spatial Ontologies instantiated for only a few organs; and, to date, it has been used to register just 48 tissue samples. However, the initial HuBMAP release demonstrates the entire workflow from human tissue acquisition to data representation within the HuBMAP Portal user interface, and includes rigorously defined imaging and data management processes.

HuBMAP has completed two of its planned eight years of development and the CCF will expand to meet the needs of new domain experts and new tissue samples by Tissue Mapping Centers and other HuBMAP funded teams that will soon increase substantially. As a result, we are sharing this early stage CCF to build awareness of the work we are doing and to obtain feedback and suggestions from the broader community, including other efforts to map the human body. In addition, we seek feedback from scientists who are interested in using HuBMAP data for their research or who may have data that would be suitable for inclusion in HuBMAP.

The CCF will need to be continually validated in terms of coverage (defined as the percent and type of human diversity that it accounts for) and quality (e.g., precision, fidelity, resolution). Coverage strongly depends on smart sampling of a diverse set of human individuals (i.e., proper coverage of the range of human sex, ethnicities, age groups, BMIs, etc.). Quality depends on the resolution of technology used, quality of 3D reconstruction of major anatomical parts and cell types, and the correctness and level-of-detail of ontology terms. The quality of existing and new datasets and workflows must be monitored to ensure new data increases CCF quality and coverage, and that the CCF design supports current and future CCF usage.

Challenges related to the CCF Ontology include the ever-expanding list of 'needed terms' such as: 1) the need to use terms across different ontologies, 2) the requirement to use a partonomy with a tree structure for better navigation in user interfaces but network graphs for realistic representation of biology, 3) the need to enter an ontology at various levels depending on the data being represented, and 4) to incorporate new changes from source ontologies into the CCF Ontology.

An ideal approach to improving the CCF over time might combine top-down expert-based (e.g., manual tissue segmentation and annotation; ontologies usage) and bottom-up, data-driven methods (e.g., machine learning applied to tissue segmentation, annotation, or registration). Manual identification of anatomical structures at macro to micro levels is usually required to generate training data for machine learning algorithms. New datasets and technologies, as well as new user needs, will both demand and make possible continuous improvements of the reference object library, ontology, and the mapping and registration processes. CCF UIs are expected to evolve to support ever more robust and detailed registration and exploration of semantically and spatially annotated tissue data.

Developing a CCF for the human body is a major undertaking that requires access to high quality and high coverage data but also human expertise across both biological domains and technological domains. It seems highly desirable to develop and agree on data formats across consortia and to develop tools and infrastructures that provide an overview and index for existing data. An example of the former is the development of ASCT+B tables across organs and experts. An example of the latter is planned work on making the RUI available to other consortia so tissue samples by other teams can be registered spatially and semantically in support of exploration via the EUI. The data and code presented here is available via the inaugural HuBMAP Portal release and GitHub repositories for EUI, RUI, and CCF Ontology [31–33].

**Acknowledgements**

We appreciate the close collaboration of MC-IU with TMC-Vanderbilt, TMC-UCSD, TMC-Florida, and NIH's NIAID over the initial two years of the HuBMAP effort. We thank Zorina S. Galis and Tyler Best, NIH for their expert comments and Bennett A. Landman, Medical-image Analysis and Statistical Interpretation (MASI) lab at Vanderbilt University Electrical Engineering and Computer Science (EECS) for providing ImageVU data and expertise.

This research has been funded by the National Institutes of Health (https://www.nih.gov) under awards OT2OD026671 [KB, EMQ, BWH, LEC, EGR, AB, JPS, MAM, GMW], OT2OD026675 [JCS], U54AI142766 [CHW, MLJ], U54DK120058 [JMS, NHP], U54HL145608 [SJ]; by the National Institute of Diabetes and Digestive and Kidney Diseases (NIDDK) Kidney Precision Medicine Project grant U2CDK114886 [KB, EMQ], and the National Institute of Allergy and Infectious Diseases (NIAID), Department of Health and Human Services under BCBB Support Services Contract HHSN316201300006W/HHSN27200002 [KMB]. The funders had no role in study design, data collection and analysis, decision to publish, or preparation of the manuscript.

**Figure Legends**

**Figure 1**. CCF Knowledge Architecture. Tissue samples and datasets are annotated using the CCF Clinical, CCF Semantic, and CCF Spatial Ontologies.

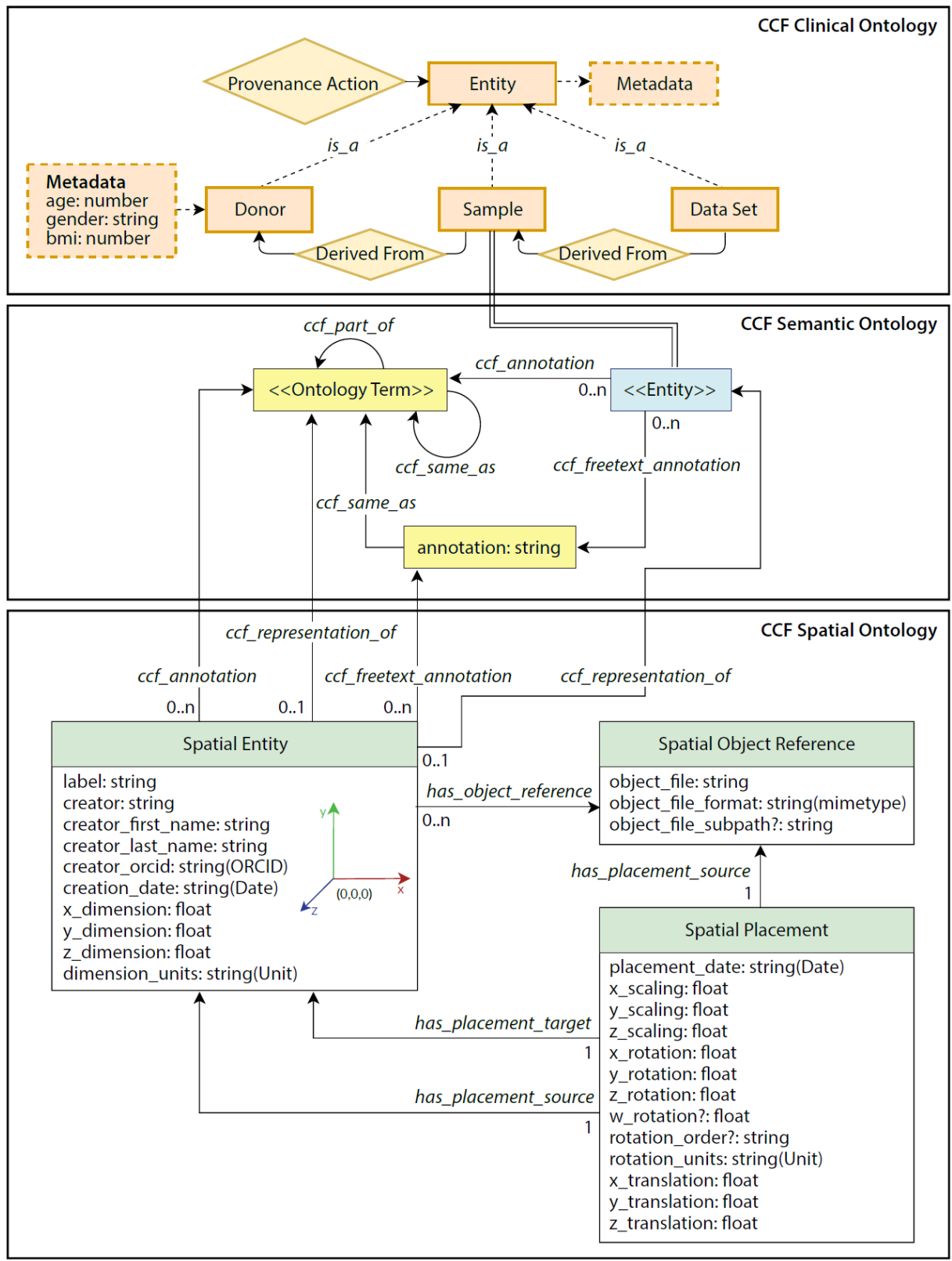

**Figure 2.** Semantic Representation of a Kidney. (a) The CCF Semantic Ontology divides the body into a set of nested named anatomical structures and cell types (the ASCT "partonomy"), from larger (left) to smaller (right) objects. (b) Construction of the CCF Semantic Ontology begins with domain experts manually developing ASCT+B tables, which indicate the most important anatomical structures (AS) and cell types (CT) for HuBMAP, organize them into a hierarchy, and map them to the 3D Reference Object Library.

**a.**

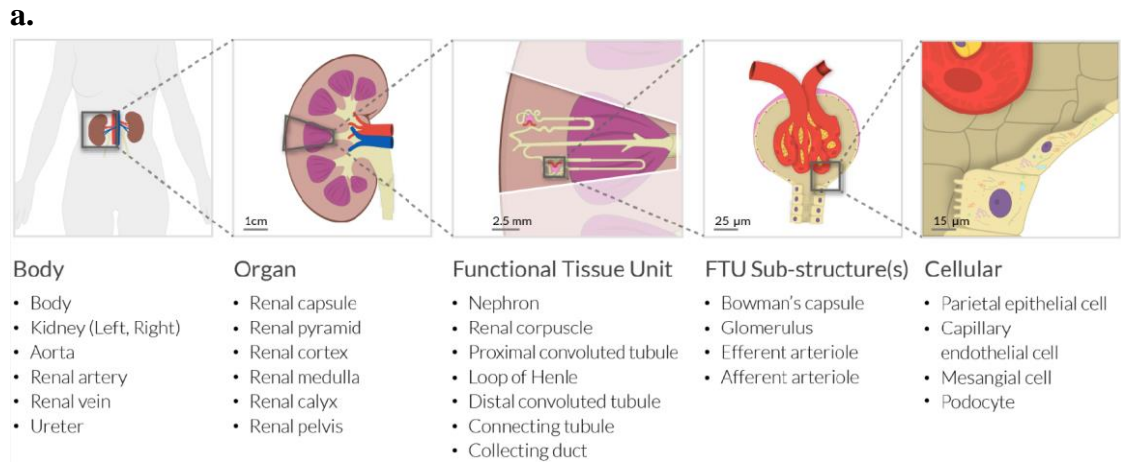

**b.**

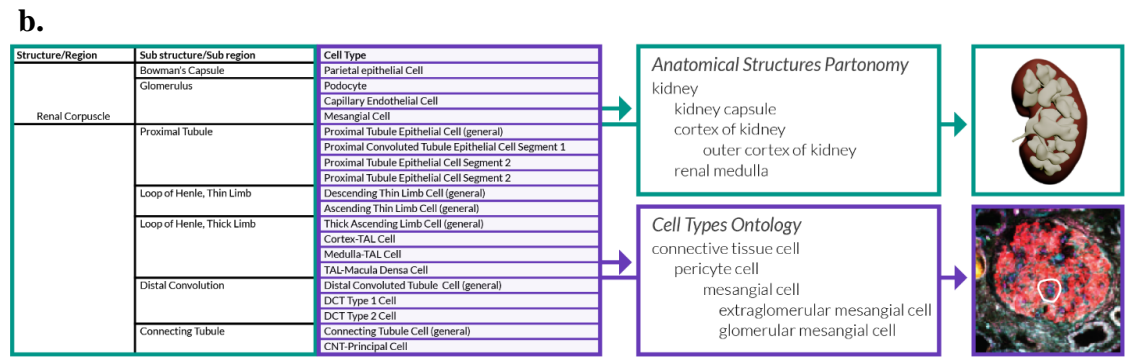

**Figure 3**. Spatial Representation of a Kidney. (a) The CCF Spatial Ontology leverages a 3D Reference Object Library to define the dimensions and shapes of ASTC entities in 3D space. (b) Construction of the CCF Spatial Ontology involves relative positioning of objects from whole body down to individual cells.

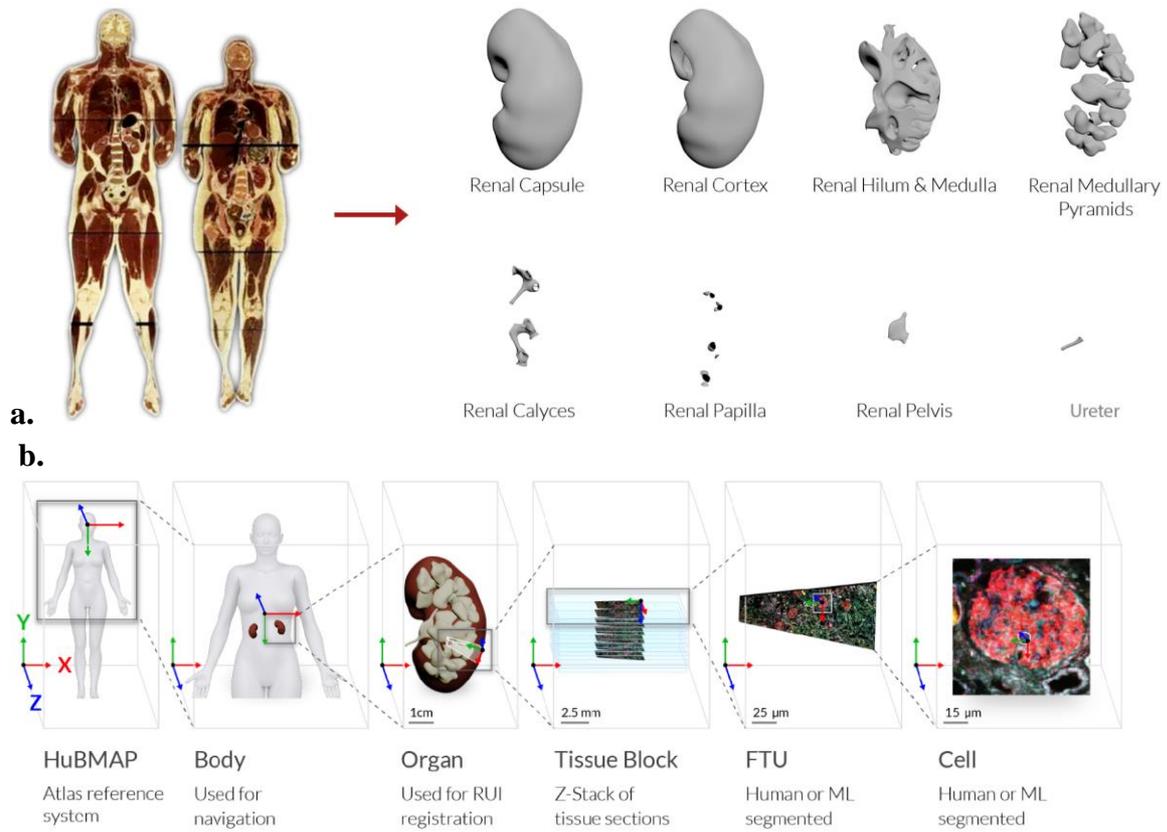

**Figure 4.** Example Instantiation of the CCF Ontologies for a Kidney Sample. Orange parts indicate CCF Clinical Ontology, gray parts the CCF Semantic Ontology, and green parts the CCF Spatial Ontology.

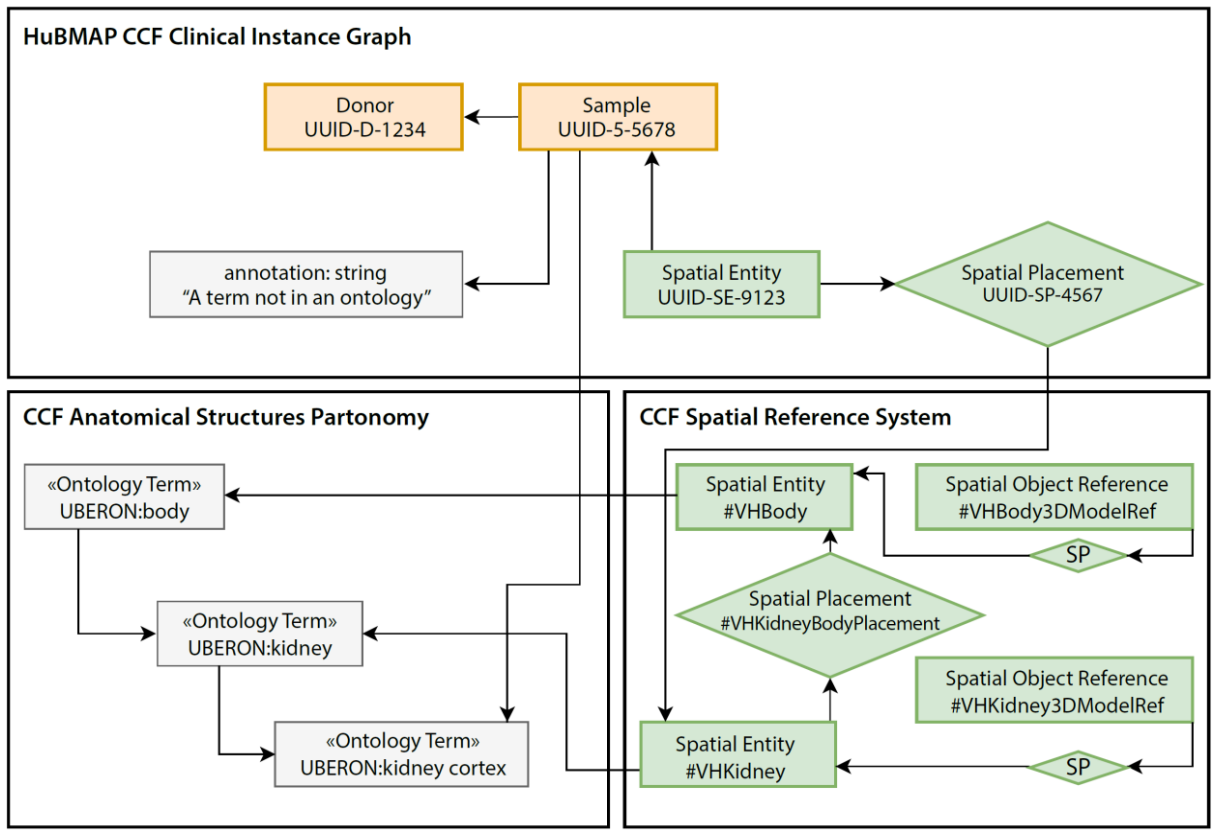

# Supplemental Information

**Contents**

## S1. CCF Semantic Ontology (anatomical structures partonomy) for the kidney

| Label (indented to indicate partonomy) | ID | Synonyms |
|---|---|---|
| body | UBERON:0013702 | |
|   abdominal cavity | UBERON:0003684 | cavity of abdominal compartment |
|     **kidney** | UBERON:0002113 | |
|       kidney capsule | UBERON:0002015 | capsula fibrosa renis |
|       cortex of kidney | UBERON:0001225 | cortex renalis |
|         outer cortex of kidney | UBERON:0002189 | kidney outer cortex |
|       renal medulla | UBERON:0000362 | kidney medulla |
|         outer medulla | UBERON:0001293 | outer renal medulla |
|         inner medulla | UBERON:0001294 | inner renal medulla |
|       renal column | UBERON:0001284 | column of Bertini |
|       renal pyramid | UBERON:0004200 | Malpighian pyramid |
|       hilum of kidney | UBERON:0008716 | hilar area of the kidney |
|       kidney interstitium | UBERON:0005215 | interstitial tissue of kidney |
|       kidney calyx | UBERON:0006517 | calices renales |
|         major calyx | UBERON:0001226 | calices renales majores |
|         minor calyx | UBERON:0001227 | calices renales minores |
|       renal pelvis | UBERON:0001224 | kidney pelvis |
|       ureter | UBERON:0000056 | |
|       renal papilla | UBERON:0001228 | kidney papilla |
|       renal fat pad | UBERON:0014464 | |
|       nephron | UBERON:0001285 | |
|         renal corpuscle | UBERON:0001229 | Malphigian corpuscle |
|           Bowman's capsule | UBERON:0001230 | Bowman's capsule |
|           glomerulus | UBERON:0000074 | renal glomeruli |
|         renal tubule | UBERON:0009773 | renal tubule (generic) |
|           proximal tubule | UBERON:0004134 | kidney proximal tubule |
|             proximal convoluted tubule | UBERON:0001287 | proximal convoluted renal tubule |
|               proximal convoluted tubule segment 1 | UBERON:0004196 | S1 portion of renal tubule |
|               proximal convoluted tubule segment 2 | UBERON:0004197 | S2 portion of renal tubule |
|             proximal straight tubule | UBERON:0001290 | S3 |
|           loop of Henle | UBERON:0001288 | Henle loop |
|             descending limb of loop of Henle | UBERON:0001289 | descending limb of Henle's loop |
|             loop of Henle ascending limb thin segment | UBERON:0004193 | ascending limb thin segment of loop of Henle |
|             thick ascending limb of loop of Henle | UBERON:0001291 | ascending thick limb |
|           distal convoluted tubule | UBERON:0001292 | distal convoluted renal tubule |
|           renal connecting tubule | UBERON:0005097 | connecting tubule |
|           collecting duct of renal tubule | UBERON:0001232 | collecting duct |

## S2. CCF Semantic Ontology (cell types partonomy) for the kidney

| Label (indented to indicate partonomy) | ID | Synonyms |
|---|---|---|
| tissue | UBERON:0000479 | |
|   epithelium | UBERON:0000483 | |
|     kidney epithelial cell | CL:0002518 | |
|       epithelial cell of nephron | CL:1000449 | |
|     meso-epithelial cell | CL:0002078 | |
|       endothelial cell | CL:0000115 | |
|         endothelial cell of vascular tree | CL:0002139 | |
|           blood vessel endothelial cell | CL:0000071 | |
|             kidney capillary endothelial cell | CL:1000892 | |
|               glomerular capillary endothelial cell | CL:1001005 | |
|               peritubular capillary endothelial cell | CL:1001033 | |
|               vasa recta cell | CL:1001036 | |
|                 vasa recta ascending limb cell | CL:1001031 | |
|                   inner medulla vasa recta ascending limb cell | CL:1001209 | |
|                   outer medulla vasa recta ascending limb cell | CL:1001210 | |
|                 vasa recta descending limb cell | CL:1001285 | |
|                   inner medulla vasa recta descending limb cell | CL:1001286 | |
|                   outer medulla vasa recta descending limb cell | CL:1001287 | |
|           endothelial cell of lymphatic vessel | CL:1000421 | |
|           endothelial cell of arteriole | CL:1000412 | |
|             kidney afferent arteriole endothelial cell | CL:1001096 | |
|             kidney efferent arteriole endothelial cell | CL:1001099 | |
|           kidney glomerular epithelial cell | CL:1000510 | |
|             epithelial cell of glomerular capsule | CL:1000450 | |
|               epithelial cell of visceral layer of glomerular capsule | CL:1000451 | |
|                 glomerular visceral epithelial cell | CL:0000653 | podocyte |
|               parietal epithelial cell | CL:1000897 | |

## S3. CCF Semantic Ontology (anatomical structures partonomy) for the spleen

| Label (indented to indicate partonomy) | ID | Synonyms |
|---|---|---|
| body | UBERON:0013702 | |
|   abdominal cavity | UBERON:0003684 | cavity of abdominal compartment |
|     **spleen** | UBERON:0002106 | |
|       spleen capsule | UBERON:0004641 | Malpighian capsule |
|       trabecula of spleen | UBERON:0001265 | spleen trabeculum |
|       spleen pulp | UBERON:1000023 | Malpighian corpuscles |
|         red pulp of spleen | UBERON:0001250 | pulpa rubra |
|           splenic cord | UBERON:0001266 | cord of Billroth |
|         white pulp of spleen | UBERON:0001959 | pulpa alba |
|           spleen lymphoid follicle | UBERON:0001249 | Malpighian body |
|             spleen primary B follicle | UBERON:0004041 | primary spleen B cell follicle |
|             spleen secondary B follicle | UBERON:0004042 | secondary spleen B cell follicle |
|               spleen germinal center | UBERON:0005196 | germinal center of spleen |
|               spleen B cell corona | UBERON:0010421 | follicle mantle |
|           periarterial lymphatic sheath | UBERON:0001960 | PALS |
|       marginal zone of spleen | UBERON:0001251 | junctional zone of spleen |
|       spleen perifollicular zone | UBERON:0005353 | |
|       hilum of spleen | UBERON:0001248 | hilum lienale |

## S4. CCF Semantic Ontology (cell types partonomy) for the spleen

| Label (indented to indicate partonomy) | ID | Synonyms |
|---|---|---|
| tissue | UBERON:0000479 | |
|   heterogeneous tissue | UBERON:0015757 | |
|     lymphomyeloid tissue | UBERON:0034769 | |
|       lymphoid tissue | UBERON:0001744 | |
|         germinal center | UBERON:0010754 | |
|           spleen germinal center | UBERON:0005196 | |

## S5. CCF Spatial Ontology summary (3D reference objects) for the kidney and spleen

Six reference objects (4 kidneys and 2 spleens) have been constructed using data from the Visible Human.

| Anatomical Structure (AS) | Female | | | Male | | |
|---|---|---|---|---|---|---|
| | Kidney, L | Kidney, R | Spleen | Kidney, L | Kidney, R | Spleen |
| Screenshot | 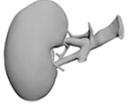 | 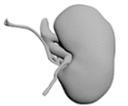 | 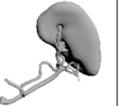 | 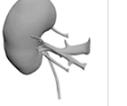 | 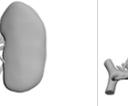 | 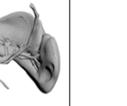 |
| #AS | 44 | 41 | 8 | 38 | 39 | 8 |
| Calyces (minor/major) | 10/4 | 10/3 | | 9/3 | 9/2 | |
| Capsule | 1 | 1 | | 1 | 1 | |
| Outer Cortex | 1 | 1 | | 1 | 1 | |
| Medulla (renal columns) | 1 | 1 | | 1 | 1 | |
| Papilla | 11 | 10 | | 9 | 10 | |
| Pelvis | 1 | 1 | | 1 | 1 | |
| Pyramids | 11 | 10 | | 9 | 10 | |
| Ureter | 1 | 1 | | 1 | 1 | |
| Hilum | 1 | 1 | 1 | 1 | 1 | 1 |
| Artery | 1 | 1 | 1 | 1 | 1 | 1 |
| Veins | 1 | 1 | 1 | 1 | 1 | 1 |
| Spleen | | | 1 | | | 1 |
| Spleen Gastric Surface | | | 1 | | | 1 |
| Spleen Colic Surface | | | 1 | | | 1 |
| Spleen Renal Surface | | | 1 | | | 1 |
| Spleen Diaphramatic Surface | | | 1 | | | 1 |
| #Triangles | 25,274,160 | 25,557,968 | 542,384 | 28,948,388 | 27,443,520 | 2,123,856 |
| #Polygons | 12,637,080 | 12,778,984 | 271,192 | 14,474,957 | 13,721,760 | 1,061,928 |
| #Vertices | 12,654,392 | 12,796,134 | 271,206 | 14,490,689 | 13,741,379 | 1,064,310 |

## S6. HuBMAP CCF Architecture

A Registration User Interface (RUI) and Exploration User Interface (EUI) leverage the CCF Semantic and Spatial Ontologies to enable users to precisely position tissue blocks within reference organs and then to use the semantic and spatial information to search, browse, and filter tissue datasets. Users access these tools through the HuBMAP Portal. The tools use information from the ASCT+B tables collected via the CCF Portal. The ASCT+B tables are used to design the CCF Semantic and Spatial Ontologies and the 3D Reference Object Library. Tissue Mapping Centers (TMCs) collect the specimens and deposit data in the IEC Data Store. 3D object collision algorithms detect where tissue blocks are placed in the RUI and automatically annotate tissue samples with terms from the CCF Semantic Ontology.

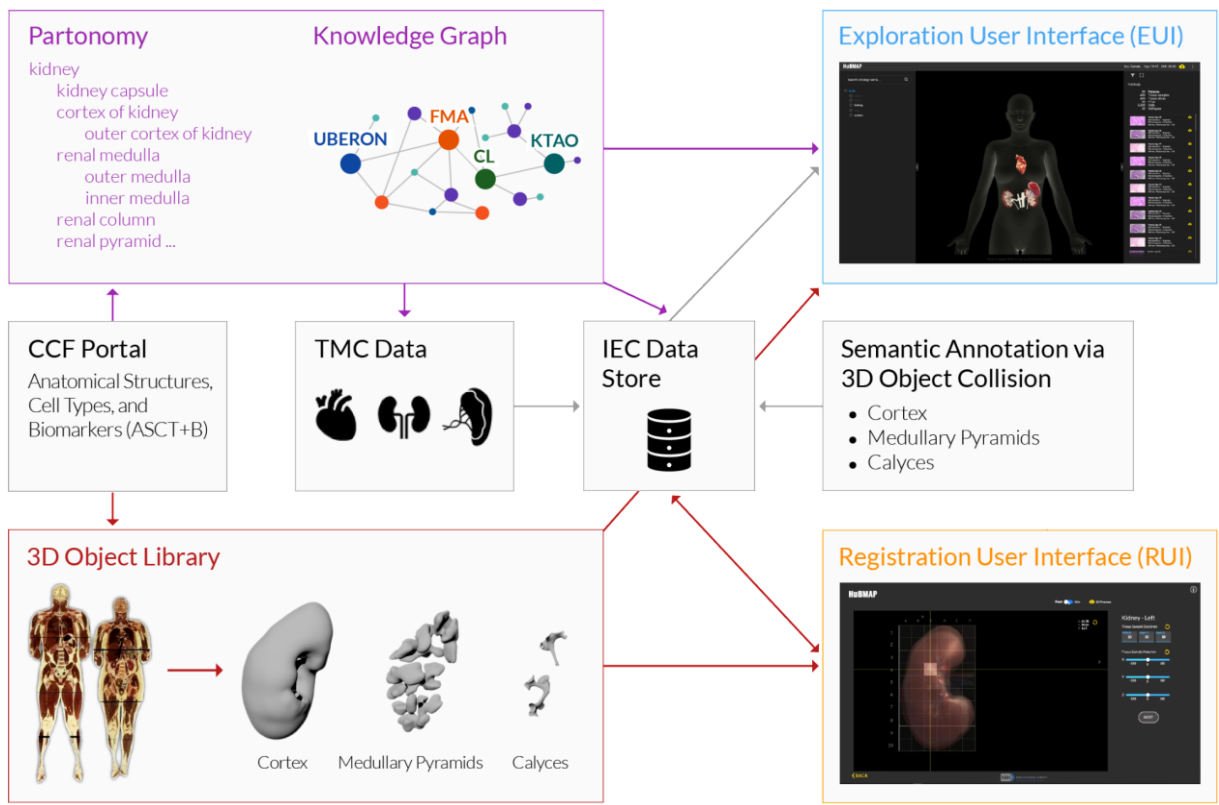

**S7. Positioning tissue blocks within organs using the Registration User Interface**

As of June 2020, a total of 24 kidney (11 blue tissue blocks in right kidney, 13 red tissue blocks in left kidney) and 24 spleen (green) tissue blocks have been registered; all registrations were confirmed with organ experts. For the spleen, there are three spleen sampling sites (top, middle, and bottom), which are further subdivided into 6 smaller blocks.

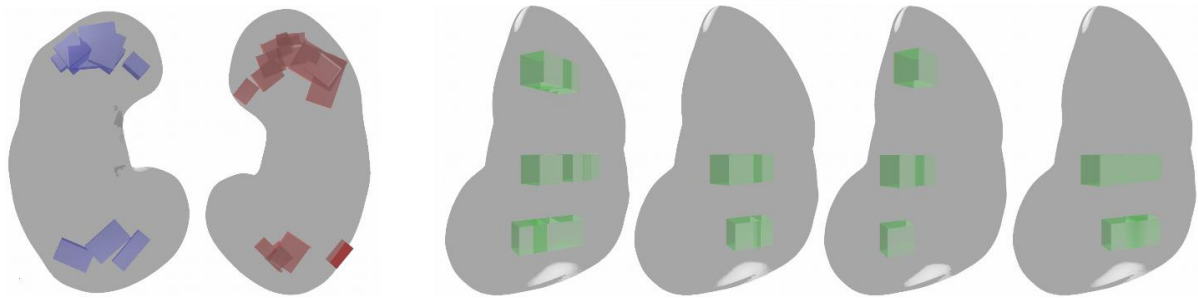